\documentclass[twocolumn,floatfix]{aastex631}
\usepackage{amssymb}
\usepackage{amsmath}
\usepackage{comment}
\usepackage{graphicx}
\usepackage{tabularx}

\usepackage{lipsum}

\shorttitle{What is the nature of the HESS J1731-347 compact object?}
\shortauthors{Sagun et al.}

\begin{document}

\title{What is the nature of the HESS J1731-347 compact object?}

\author[0000-0001-5854-1617]{Violetta Sagun}
\affiliation{CFisUC, Department of Physics, University of Coimbra, Rua Larga P-3004-516, Coimbra, Portugal}

\author[0000-0001-9545-466X]{Edoardo Giangrandi}
\affiliation{CFisUC, Department of Physics, University of Coimbra, Rua Larga P-3004-516, Coimbra, Portugal}
\affiliation{Institut für Physik und Astronomie, Universität Potsdam, Haus 28,
Karl-Liebknecht-Str. 24/25, Potsdam, Germany}

\author[0000-0003-2374-307X]{Tim Dietrich}
\affiliation{Institut für Physik und Astronomie, Universität Potsdam, Haus 28,
Karl-Liebknecht-Str. 24/25, Potsdam, Germany}
\affiliation{Max Planck Institute for Gravitational Physics (Albert Einstein Institute), Am Mühlenberg 1, Potsdam 14476, Germany}

\author[0000-0002-4947-8721]{Oleksii Ivanytskyi}
\affiliation{Incubator of Scientific Excellence---Centre for Simulations of Superdense
Fluids, University of Wrocław, 50-204, Wroclaw, Poland}

\author[0000-0000-0000-0000]{Rodrigo Negreiros}
\affiliation{Instituto de Física, Universidade Federal Fluminense–UFF, Niterói-RJ, 24210-346, Brasil}

\author[0000-0001-6464-8023]{Constança Providência}
\affiliation{CFisUC, Department of Physics, University of Coimbra, Rua Larga P-3004-516, Coimbra, Portugal}

\begin{abstract}
Once further confirmed in future analyses, the radius and mass measurement of HESS J1731-347 with $M=0.77^{+0.20}_{-0.17}~M_{\odot}$ and $R=10.4^{+0.86}_{-0.78}~\rm km$ will be among the lightest and smallest compact objects ever detected. This raises many questions about its nature and opens up the window for different theories to explain such a measurement. In this article, we use the information from~\citet{Doroshenko:2022} on the mass, radius, and surface temperature together with the multimessenger observations of neutron stars to investigate the possibility that HESS J1731-347 is one of the lightest observed neutron star, a strange quark star, a hybrid star with an early deconfinement phase transition, or a dark matter-admixed neutron star. The nucleonic and quark matter are modeled within realistic equation of states (EOSs) with a self-consistent calculation of the pairing gaps in quark matter. By performing the joint analysis of the thermal evolution and mass-radius constraint, we find evidence that within a 1$\sigma$ confidence level, HESS J1731-347 is consistent with the neutron star scenario with the soft EOS as well as with a strange and hybrid star with the early deconfinement phase transition with a strong quark pairing and neutron star admixed with dark matter.

\end{abstract}


\section{Introduction} 
\label{sec:intro}

One open question of modern physics is it to find out if strongly interacting matter at high densities undergoes a phase transition to deconfined quarks and gluons. To tackle this problem, many fields of physics have been working together, including gravitational wave (GW) and multimessenger astrophysics, nuclear physics, and high-energy physics. While the latter probes finite temperature regimes, astrophysical observations of compact stars (CSs) give mainly access to vanishing temperature and high baryon density regimes, which cannot be probed with terrestrial experiments. In fact, recent detections of binary neutron star (NS) and NS-black hole mergers~\citep{Abbott_2018,LIGOScientific:2020aai} have opened a way to constrain zero-temperature NS matter properties during the inspiral phase, whereas the next generation of the GW telescopes is planned to reach enough sensitivity to relate it to the finite temperature equation of state (EOS) during the merger and postmerger phases ~\citep{Raithel:2022efm}.

Considering the existing observational data of the heaviest known NSs~\citep{PSRj03480432Article, Fonseca:2021wxt, Romani:2021xmb, Romani:2022jhd}, the EOS of strongly interacting matter at densities above twice the normal saturation density ($2n_{0}$) is required to be stiff leading to typical NS radius measurements of $\sim 11-14$~km at $M \gtrsim 1.4~M_{\odot}$. Therefore, despite being in agreement with theoretical calculations for the minimum mass of NSs, e.g., $M=0.88-1.28~M_{\odot}$~\citep{Strobel:2000mg, Lattimer:2000nx}, the recently announced measurement of HESS J1731-347 with $M=0.77^{+0.20}_{-0.17}~M_{\odot}$ and $R=10.4^{+0.86}_{-0.78}$ km~\citep{Doroshenko:2022} challenges our understanding of the EOS at densities $1-2~n_{0}$. In fact, simulations of supernova explosions predict the gravitational mass of the lightest possible NS to be $M=1.17~M_{\odot}$, corresponding to a baryonic mass of $M=1.25-1.31M_{\odot}$~\citep{Suwa:2018uni}.

The HESS J1731-347 measurement is also interesting as the first simultaneous measurement of the mass, radius, and surface temperature of a CS and opens the possibility to study its thermal evolution. The estimated $M=0.77^{+0.20}_{-0.17}~M_{\odot}$ relies on the fact that the object has a uniform-temperature carbon atmosphere and that the star is located at a distance of $2.5\ \rm kpc$. Further studies are needed to reduce the uncertainties and understand the validity of the obtained results~\citep{Alford:2023waw}.

It was proposed that HESS J1731-347 could be a candidate for a strange quark star (QS) comprising all existing mass-radius measurements of CSs~\citep{DiClemente:2022wqp, Horvath:2023uwl, Das:2023qej}. However, unpaired quark matter undergoes fast cooling, leading to difficulties in reproducing the surface temperature, presented by~\cite{Doroshenko:2022}. In general, the cooling of a CS is defined by the internal composition, which is determined by the processes that operate inside the star~\citep{Potekhin:2015qsa}. The fastest cooling direct Urca (DU) process corresponds to $\beta$- and inverse $\beta$-decay of neutron and $d$-quark in nuclear and quark matter, respectively. The nucleonic DU process has a threshold controlled by the number density of electrons according to the momentum conservation. As soon as the DU process is on, it leads to an intense neutrino emission and a consequent rapid drop in the surface temperature. On the other hand, the quark DU threshold condition is much easier to satisfy. In fact, such a fast cooling of strange stars contradicts the estimated redshifted surface temperature of HESS J1731-347 compact object $T_{s}^{\infty} =2.05^{+0.09}_{-0.06}$ MK~\citep{Doroshenko:2022} itself being rather high as for the age of $\sim 27$ kyr~\citep{Beznogov:2014yia}. This difficulty could be overcome within the strange QS or hybrid star (HS) with an early quark deconfinement~\citep{Ivanytskyi:2022oxv,Ivanytskyi:2022mlk, Brodie:2023pjw} scenarios, where strong quark pairing leads to color superconductivity and strongly suppresses cooling of the quark core~\citep{Schaab:1997hx,Weber:2004kj}, while the baryonic matter (BM) provides a moderate cooling. The possible suppression of the nucleonic cooling is related to the superfluidity of neutrons and superconductivity of protons via the Cooper pairs breaking and formation (PBF)~\citep{Yakovlev:2000jp}.

The questions at what density the deconfinement phase transition occurs and what the signals of the quark matter formation are still open. The elliptic flow in heavy-ion collisions and a combined analysis of multimessenger constraints~\citep{Annala:2021gom,Huth:2021bsp} suggest that strongly interacting matter softens at a high density, which might correspond to a phase transition to quark-gluon plasma. 

In this work, we examine the nature of HESS J1731-347 by considering how the simultaneously measured mass, radius, and surface temperature agree with the present theoretical understanding of the properties of strongly interacting matter and color superconductivity at high densities. 

We also discuss an alternative origin of HESS J1731-347 as a dark matter (DM)-admixed NS, a scenario that gained a lot of attention recently ~\citep{Goldman:2013qla, Tolos:2015qra,2018PhRvD..97l3007E,2019JCAP...07..012N, Ivanytskyi:2019wxd,2020MNRAS.495.4893D,DiGiovanni:2021ejn,Sagun:2021oml, Dengler:2021qcq, Leung:2022wcf,Rafiei_Karkevandi_2022}. In fact, DM could be accumulated in the core of a NS leading to a decrease of the total gravitational mass, radius, and tidal deformability, which we will perceive as the effect similar to softening of the EOS~\citep{Giangrandi:2022wht}. At the same time, this scenario of asymmetric noninteracting DM agrees with cosmological and astrophysical observations, e.g. the Bullet Cluster~\citep{Randall:2008ppe}, and provides a description of HESS J1731-347.

The article is organized as follows. In Sections~\ref{sec:NS},~\ref{sec:SS} and~\ref{sec:HS}, we present the NS, strange QS, and HS scenarios, respectively. A detailed description of the thermal evolution of CSs is presented in Appendix~\ref{app:cooling}. The possibility of HESS J1731-347 to be a DM-admixed NS with a 1$\sigma$ confidence interval (CI) is studied in Section \ref{sec:DMANS} with a detailed explanation in Appendix~\ref{app:DM}. Section~\ref{sec:Concl} summarizes the results.

\section{Source Classes}

\subsection{Neutron star} 
\label{sec:NS}

\begin{figure}[t!]
\centering
\includegraphics[width=0.95\columnwidth]{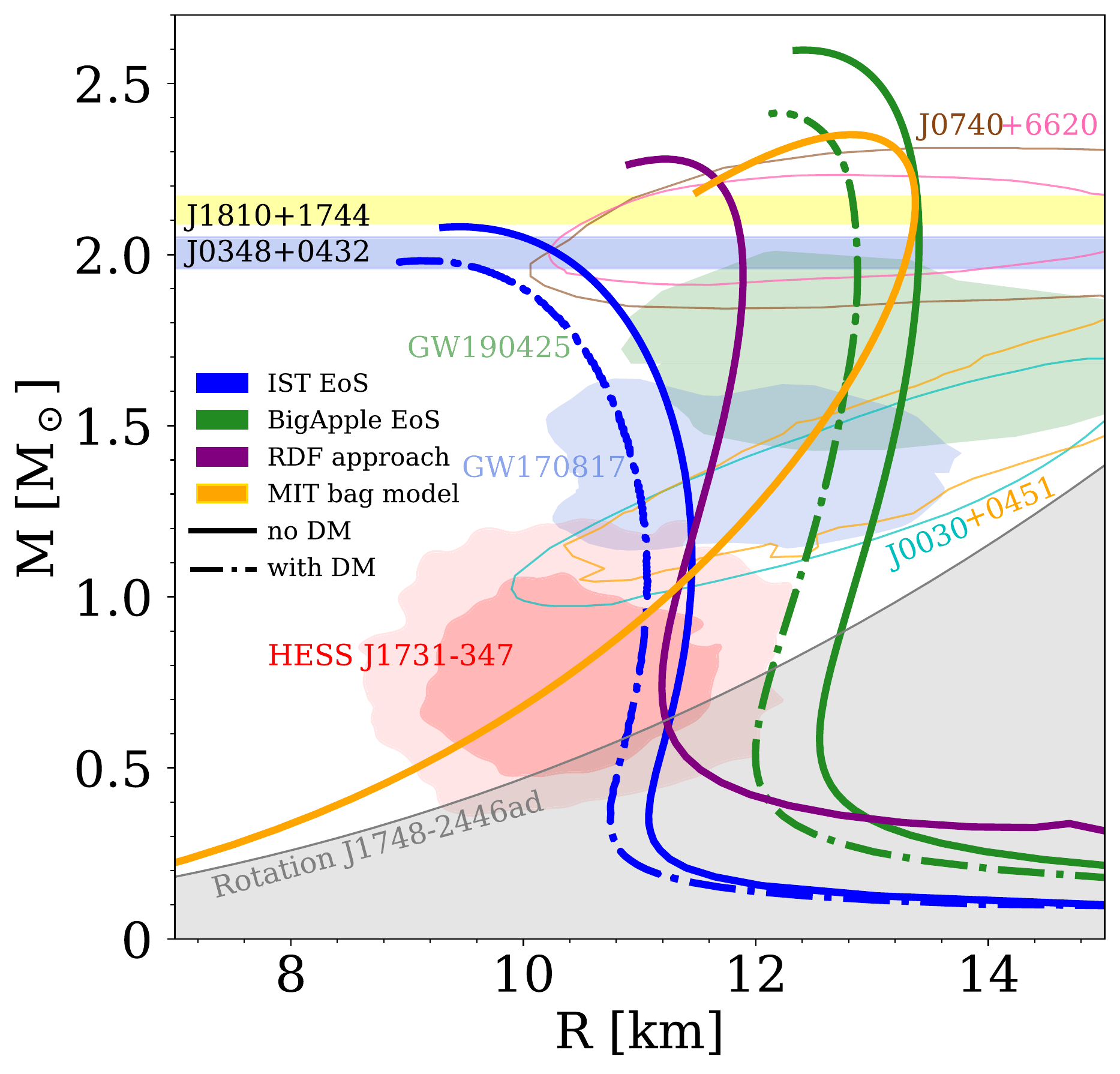}
\caption{
Mass-radius relations obtained within the scenarios of purely hadronic NS with soft (blue solid curve) and stiff (green solid curve) EOSs, strange QS (orange solid curve), and HS (purple solid curve), while the scenario of purely hadronic NS admixed with DM is represented by dash-dotted curves. The corresponding EOSs are mentioned in the legend. Light blue and yellow bands represent 1$\sigma$ constraints on the mass of PSR J0348+0432~\citep{PSRj03480432Article} and PSR J1810+1744~\citep{Romani:2021xmb}. Orange and light blue contours show the NICER measurement of PSR J0030+0451~\citep{Riley:2019yda,Miller_2019}, while pink and brown contours depict the PSR J0740+6620 measurement~\citep{Miller:2021qha,Riley:2021pdl}. LIGO-Virgo observations of GW170817~\citep{Abbott_2018} and GW190425~\citep{LIGOScientific:2020aai} binary NS mergers are shown in royal blue and green. The 1$\sigma$ and 2$\sigma$ contours of HESS J1731-347~\citep{Doroshenko:2022} are plotted in dark and light red. The shaded region is forbidden by the rotation of the fastest spinning pulsar PSR J1748-2446ad~\citep{Hessels:2006ze}. 
\label{fig:MR}}
\end{figure}

Modeling the internal structure of HESS J1731-347 requires consistency with the properties of the nuclear matter ground state, chiral effective field theory~\citep{Tews:2012fj}, existing constraints on the mass-radius relation~\citep{Miller_2019,Riley:2019yda,Raaijmakers:2019dks,Miller:2021qha,Riley:2021pdl,PSRj03480432Article, Fonseca:2021wxt, Romani:2021xmb, Romani:2022jhd} and tidal deformability of NSs~\citep{Abbott_2018, LIGOScientific:2020aai}. 
The scenario of purely hadronic matter described with an EOS, which respects the above requirements, suggests a minimal assumption about the HESS J1731-347 nature. At the densities expected inside an NS of subsolar mass such a nuclear EOS can be strongly constrained by the microscopic Brueckner-Hartree-Fock calculations based on realistic nuclear potentials fitted to the nuclear scattering data~\citep{Yamamoto:2015lwa,Yamamoto:2017wre}. However, the uncertainty of nuclear EOS obtained with these methods becomes important for modeling NSs heavier than $0.5~M_\odot$. Therefore, due to the exploratory reasons in this article, we prefer to consider the possibilities of soft and stiff nuclear EOSs instead of relying on the results of the microscopic calculations.
For this purpose, we utilize the set B of the induced surface tension (IST)~\citep{NSOscillationsEOS} and BigApple~\citep{Fattoyev:2020cws} EOSs, respectively. As is seen from Fig.~\ref{fig:MR}, while stiff hadronic EOS is completely outside the $2\sigma$ CI, the soft one is able to fit HESS J1731-347 constraint within the $1\sigma$ CI.

The recently reported data on the thermal evolution HESS J1731-347 suggest its slow cooling~\citep{Doroshenko:2022}.
To reproduce these data we consider $^{1}S_{0}$ and $^{3}P_{2}$ neutron superfluidity and $^{1}S_{0}$ proton superconductivity described by the SFB model~\citep{Schwenk:2002fq}, phenomenological gap obtained from the fit of Cas~A~\citep{Shternin:2010qi} and CCDK~\citep{Chen:1993bam} model, respectively (see Appendix~\ref{app:cooling}).

\begin{figure}[ht!]
\includegraphics[width=0.95\columnwidth]{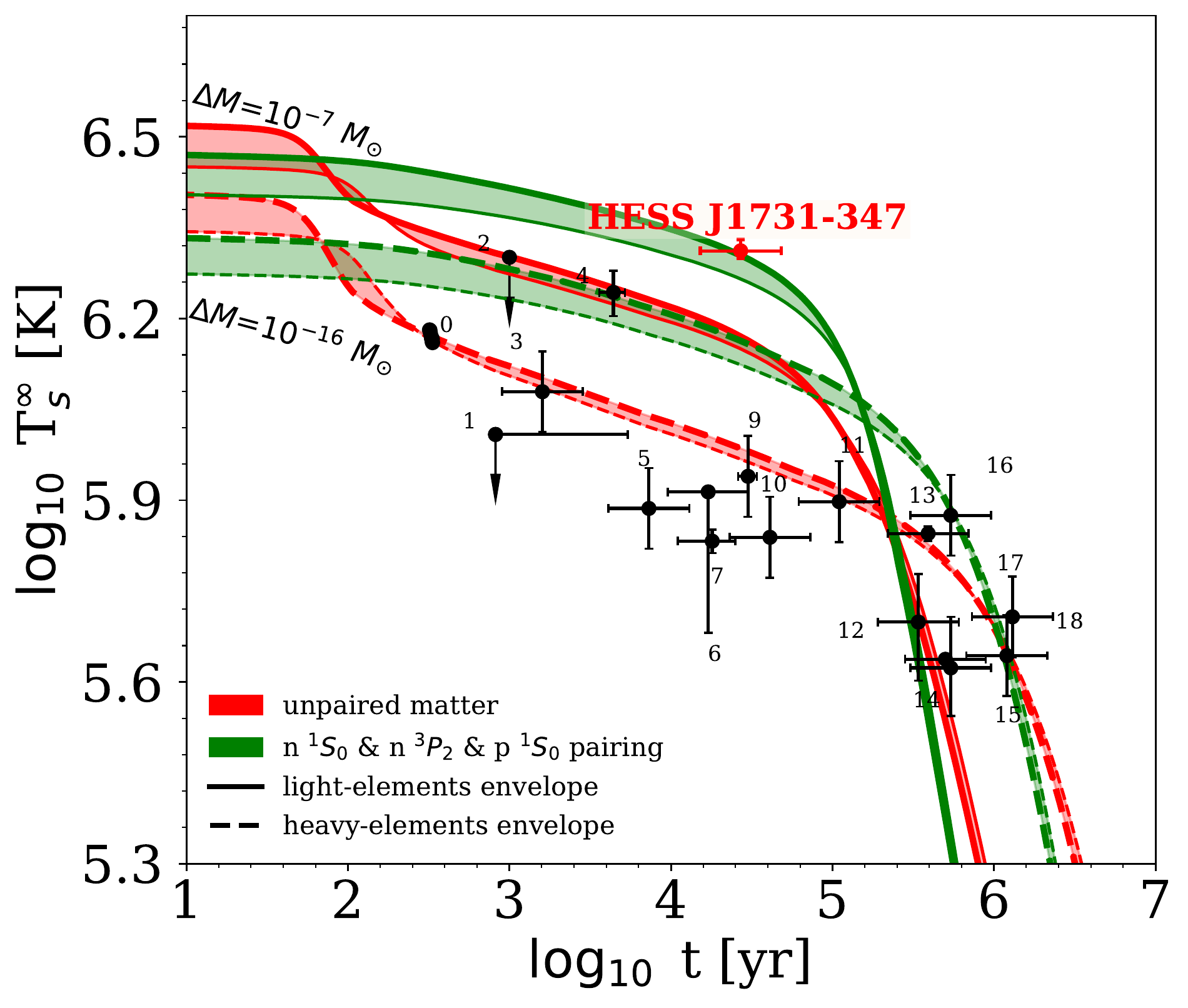}
\includegraphics[width=0.95\columnwidth]{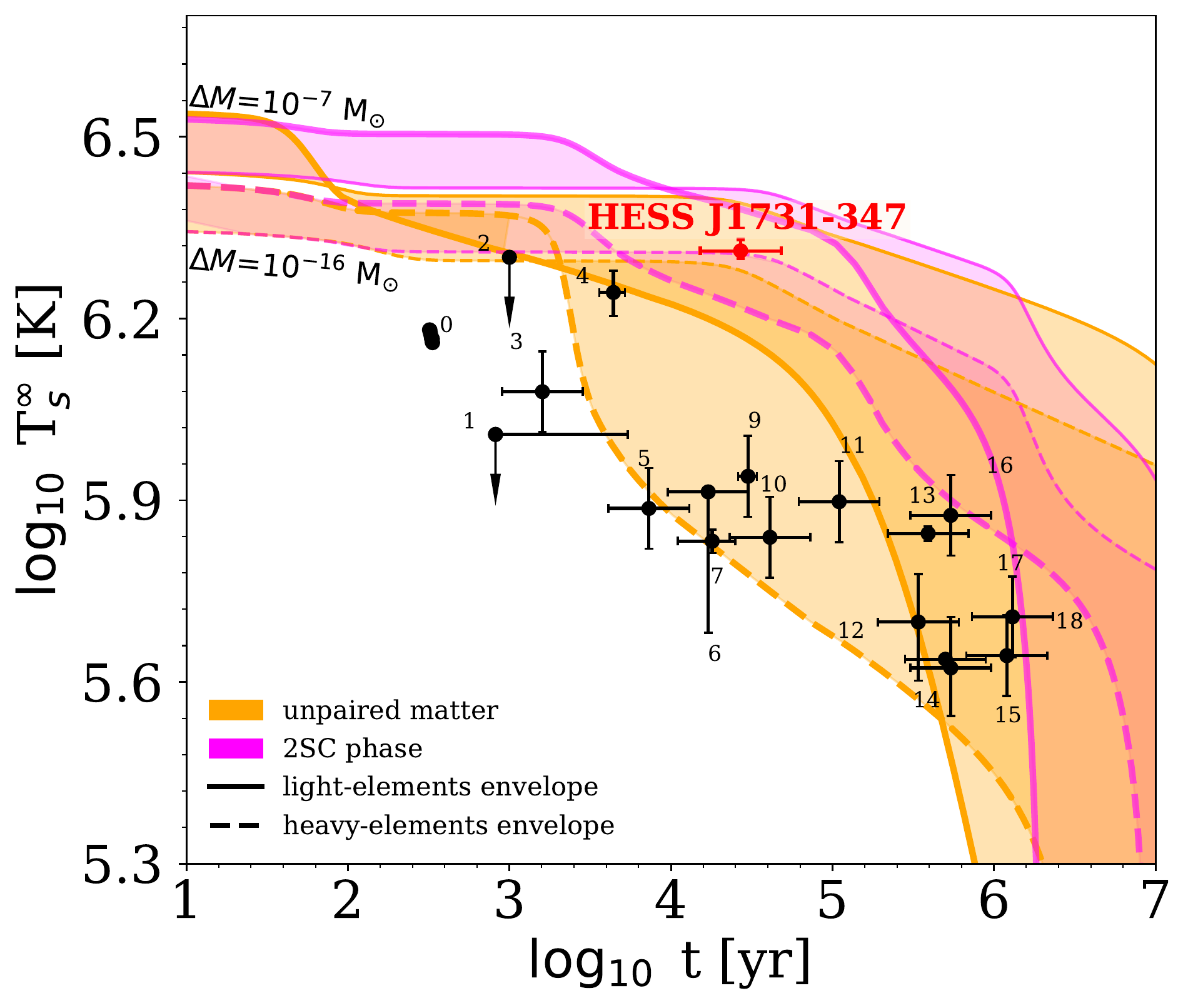}
\caption{Cooling curves for NSs (upper panel) and HSs (lower panel) of $0.77^{+0.20}_{-0.17}~M_{\odot}$ for unpaired and paired matter. $T^{\infty}_{S}$ represents the surface temperature at infinity. \textbf{Upper panel}: the red bands correspond to the unpaired NS matter, while paired n and p in the $^{1}S_{0}$ channel and n in the $^{3}P_{2}$ channel are shown in green. The color bands represent the mass measurement in the 1$\sigma$ CI, whereas the thick and thin lines define the stars of  $0.97~M_{\odot}$ and $0.6~M_{\odot}$, respectively. \textbf{Lower panel}: the thermal evolution of HSs with unpaired and paired quark matter are depicted by the magenta and orange bands. The solid and dashed curves correspond to the light-elements and heavy-elements envelopes, respectively. The data for Cas A (number 0) and the rest of the points are taken from~\cite{Wijngaarden:2019tht} and~\cite{Beznogov:2014yia}, respectively. An updated surface temperature for HESS J1731-347 is taken from~\cite{Doroshenko:2022}, while its age is considered from~\cite{Beznogov:2014yia} (see details in Appendix \ref{app:cooling}).
\label{fig:cooling}}
\end{figure}

As is seen from the upper panel of Fig.~\ref{fig:cooling}, the scenario of soft hadronic matter with paired nucleons and light-elements envelope (the green band with solid curves) is consistent with the observational data on the thermal evolution of HESS J1731-347.

\subsection{Quark star} 
\label{sec:SS}

The small mass and radius of the HESS J1731-347 object assume its gravitational binding energy to be large, which can be provided by the scenario of a strange QS. The simplest description of the strange quark matter corresponds to the MIT bag model-like EOS, which relates the stellar matter pressure $p$ and energy density $\varepsilon$ via the relation $p=\varepsilon/3-4B/3$ ~\citep{Chodos:1974pn,Alcock:1986hz}.  The orange solid curve in Fig.~\ref{fig:MR} passes through the point $M=0.77~M_\odot$, $R=10.4$ km and is obtained for the central value of the bag constant range $B^{1/4}=134^{+12}_{-11}$ MeV. This range of $B$ is obtained by fitting the HESS J1731-347 mass-radius constraint within the $1\sigma$ CI with the present EOS. 

\subsection{Hybrid star} 
\label{sec:HS}

The positive bag constant phenomenologically models quark confinement at small densities, when quark matter with negative pressure is dynamically unstable against conversion to hadronic matter with $p>0$. This assumes the existence of a hadron envelope enclosing the quark core of an NS and motivates us to consider HESS J1731-347 as a hybrid quark-hadron object. For this we utilize the hybrid EOS developed by~\citet{Ivanytskyi:2022oxv}. Its quark part is based on a chirally symmetric relativistic density functional (RDF) approach for two-flavor color superconducting (2SC) quark matter. Within this RDF approach, quark confinement is phenomenologically modeled by the fast growth of the quark quasiparticle self-energy in the confining region, where the quark EOS is matched to the DD2npY-T hadronic one~\citep{Shahrbaf:2022upc} by means of the Maxwell construction. While most of the RDF approach parameters are fitted to vacuum phenomenology of QCD, values of the dimensionless couplings controlling strength of the vector repulsion between quarks $\eta_V=0.265$ and diquark pairing $\eta_D=0.555$ were chosen by~\citet{Ivanytskyi:2022oxv} in order to provide the best agreement with the observational constraints on the NS mass-radius diagram shown in Fig.~\ref{fig:MR} and on a $1.4~M_\odot$ NS tidal deformability extracted from the GW170817 GW signal~\citep{Abbott_2018}. As is seen from Fig.~\ref{fig:MR}, this parameterization of a hybrid quark-hadron EOS agrees with the HESS J1731-347 mass-radius constraint within the $1\sigma$ CI. Furthermore, strong quark pairing suppresses cooling of the 2SC quark matter making the HS scenario also consistent with the data on the thermal evolution of HESS J1731-347 (see the lower panel of Fig.~\ref{fig:cooling}).


\subsection{Dark matter admixed neutron star} 
\label{sec:DMANS}

We model DM as a relativistic Fermi gas of noninteracting particles with spin one-half and mass $m_{DM}$ accumulated inside NSs with a relative fraction $f_{DM}$. For more details about the DM EOS and two-fluid approach, see Appendix~\ref{app:DM}. The dash-dotted curves in Fig.~\ref{fig:MR} show the effect of DM particles with $m_{DM}=2.8$ GeV and $f_{DM}=4.75\%$ accumulated in the core of NSs. To exclude the uncertainties of the underlying BM EOS from consideration the scan is performed for the soft IST EOS (blue curve)~\citep{NSOscillationsEOS} and stiffer BigApple EOS (green curve)~\citep{Fattoyev:2020cws}. The values of the DM particle's mass and relative fraction were chosen to provide an agreement with the 2$\sigma$ constraints of HESS J1731-347 for both BM EOSs. 

\begin{figure}[ht!]
\centering
\includegraphics[width=0.95\columnwidth]{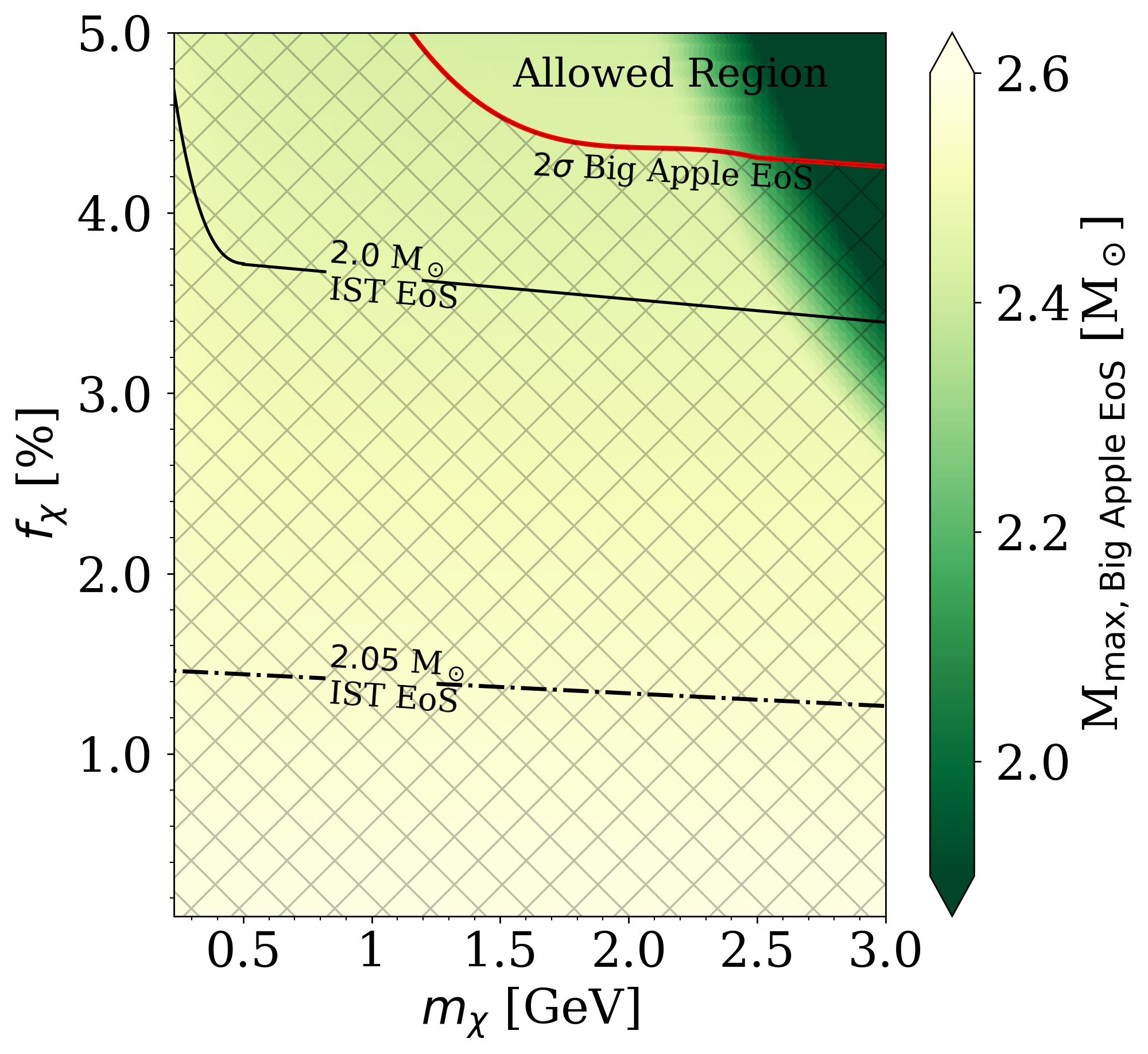}
\caption{Scan over the particle's mass $m_\mathrm{DM}$ and fraction $f_\mathrm{DM}$ of DM that reproduce M and R measurement of HESS J1731-347. While the colormap represents the total maximum gravitational mass of DM-admixed NSs using the BigApple EOS, the contour lines represent the IST EOS configurations. Dash-dotted and solid black curves are the contour lines showing the maximum mass obtainable with the IST EOS. The red solid line shows the limit at which both stiff and soft EOSs fulfill the 2$\sigma$ constraint of HESS J1731-347. Above which, in the upper right corner of the plot, there is the allowed region where both baryonic models fulfill the HESS constraint.  
\label{fig:scan}}
\end{figure}

As can be seen, the presence of a dense DM core leads to a strong reduction of the total gravitational mass and radius of
stars, which resembles a softening of the BM EOS. This degeneracy between the effect of DM and possible change of the strongly interacting matter properties at high density was studied in~\citet{Ivanytskyi:2019wxd,Sagun:2021oml,Giangrandi:2022wht}. In fact, the scan over mass and fraction of DM in Fig.~\ref{fig:scan} shows that the same effect could be seen for heavier DM particles and smaller fractions. The color map represents the total maximum gravitational mass of DM-admixed NSs for the BigApple EOS. The overlap between the two scans yields the allowed region of mass and fraction of DM (above the red curve in Fig.~\ref{fig:scan}) that reproduces M and R measurement of HESS J1731-347 (2$\sigma$ CI), free from the BM EOS uncertainties. 

Moreover, the DM-admixed NS scenario remains the thermal evolution unaffected, as DM, within the considered candidate, does not take part in cooling. Effectively, the thermal evolution of a NS admixed with asymmetric DM interacting only through gravity with BM is equivalent to cooling of a pure NS of a smaller mass~\citep{Avila:2023rzj}. Thus, this scenario is consistent with the HESS J1731-347 data.

\section{Conclusions}
\label{sec:Concl}

We analyzed four different scenarios of the possible internal composition of a central compact object within the supernova remnant HESS J1731-347 in light of the recent measurement presented by~\citet{Doroshenko:2022}. We discuss that the soft nucleonic EOS is able to simultaneously reproduce the mass and radius measurements within the 1$\sigma$ CI, and give a good description of the surface temperature. This scenario also includes the possibility of the soft hadronic EOS at 1-2$n_{0}$, while at $n\geq 2n_{0}$ the EOS could attain an extra stiffening due to, e.g. density dependent repulsion between the constituents. Such behavior is also consistent with the proton flow constraint~\citep{Danielewicz:2002pu}. From the analysis of various models of pairing gaps, we conclude that a combination of n and p singlet pairing with n triplet pairing described within the SFB and CCDK models, and phenomenological gap obtained from the fit of Cas~A, respectively, with light-elements envelope provide the best fit of the data. 

HESS J1731-347 could also be explained as an HS with an early deconfinement phase transition that occurs below twice nuclear saturation density. Thus, HS would contain a big quark-gluon plasma core, which could potentially lead to rapid cooling related to an operating quark DU process. However, a self-consistent derivation of the quark pairing gap and RDF model formulation makes us conclude that quarks exist in the 2SC phase that suppresses rapid cooling and provides an agreement with the surface temperature measurement.

Moreover, the strange QS scenario can also reproduce mass, radius, and surface temperature very well, as, similar to the HS scenario, paired quarks suppress neutrino emission, while, in comparison to the HS case, the photon emission from the surface will be even lower due to the smaller star's radius. However, the three abovementioned scenarios are in tension with the recent supernova simulations that predict the lowest compact star to be 1.17$M_\odot$.

As an alternative scenario, we considered HESS J1731-347 to be a NS admixed with DM, which results in the effective softening of the EOS and creation of more compact configurations. This scenario leaves the thermal evolution unaffected, as asymmetric noninteracting DM interacts only through gravity with BM. Based on the performed scan over model parameters we found that fermionic DM particles with mass above 1.15 GeV and fraction above 4.2$\%$ provide a full agreement with HESS J1731-347 2$\sigma$ CI measurement for both stiff and soft baryonic EOSs. The analysis was made for two different EOSs that cover the parameter range to exclude the BM uncertainties from consideration. The performed scan over mass and fraction of DM shows that the same effect could be seen by increasing the DM particle's mass and decreasing its fraction. 

We argue that in comparison to GW170817, GW190425, NICER, and heaviest CSs measurements that probe the properties of strongly interacting matter at high densities, HESS J1731-347 provides an important piece of information in the range of 1-2 nuclear saturation density. 

Future observations of the HESS J1731-347 object are required, as well as the study of the impact of different effects, e.g., existence of the possible hot/cold spots on the surface of the star, atmosphere composition, distance to the object, etc.  While the low mass and radius measurement is confirmed, it will put the most stringent constraint on the strongly interacting matter at 1-2$n_{0}$ density range and possible DM-rich environment around the star.

\begin{acknowledgments}
The work of E.G., C.P., and V.S. was supported by national funds from FCT – Fundação para a Ciência e a Tecnologia, I.P., within the Projects No.  UIDB/04564/2020, UIDP/04564/2020, EXPL/FIS-AST/0735/2021. E.G. also acknowledges the support from Project No. PRT/BD/152267/2021. C.P. is supported by Project No. PTDC/FIS-AST/28920/2017. The work of O.I. was supported by the program Excellence Initiative--Research University of the University of Wrocław of the Ministry of Education and Science. R.N. acknowledges financial support from CAPES, CNPq, and FAPERJ. This work is part of the project INCT-FNA Proc. No. 464898/2014-5 as well as FAPERJ JCNE Proc. No. E-26/201.432/2021.
\end{acknowledgments}

\vspace{5mm}

\appendix

\section{Thermal evolution of compact stars}
\label{app:cooling}

Born in supernova explosions or through the coalescence of light CSs, NSs cool down through a combination of neutrino emission from their interior and thermal radiation from the surface. The former process is directly determined by the internal composition of a star. Low- and medium-mass stars usually cool down through so-called slow and intermediate cooling processes, i.e. modified Urca, bremsstrahlung, and PBF~\citep{Page:2005fq}. 

Once the relative abundances of involved baryonic and leptonic species are high enough, the DU process starts to operate leading to rapid cooling. It corresponds to the $\beta$- and inverse $\beta$-decay that operate after the triangle inequality of Fermi momenta $p_{F,i} + p_{F,j} \geq p_{F,k}$ is satisfied. Accounting for charge neutrality and the relation between the Fermi momenta and the number density of each particle, we obtain the DU threshold corresponding to the minimal proton fraction of $\sim 11\%$ of the total baryon density~\citep{Lattimer:1991ib}. In quark matter the threshold for the DU reactions, $d \rightarrow u + e^{-} +  \bar{\nu}_{e}$ and $u + e^{-} \rightarrow  d + \nu_{e}$, is much easier to satisfy leading to an emissivity of $\epsilon \sim T^{6}$. For comparison, the modified Urca processes give only $\epsilon \sim T^{8}$~\citep{Iwamoto:1980eb, Iwamoto:1982zz}. 

At vanishing temperature, an attraction between nucleons or quarks leads to the existence of pairs, where the particle excitations are gapped and the cooling mechanism is drastically suppressed. At temperatures below the critical temperature of nuclear superfluidity, $T \ll T_{c}$, the neutrino emission is suppressed by a Boltzman factor $e^{\frac{-\Delta}{T}}$, where $\Delta$ is the energy gap. At $T_{c}$, the effect of PBF results in neutrino emissivity of $\epsilon \sim T^{7}$~\citep{Page:2005fq}.

In quark matter depending on the abundance of strange quarks, it is possible to distinguish the color-flavor-locked (CFL) phase, in which the quarks form Cooper pairs, whose color properties are correlated with their flavor properties in a one-to-one correspondence between three color pairs and three flavor pairs, and the two-color superconducting (2SC) phase, characterized by the absence of the s-quark and the appearance of u-d diquark condensate in a selected direction in color space~\citep{Alford:2009qj}. In fact, the quark pairing could be more diverse, e.g. gapless 2SC, crystalline CSC, gapless CFL (gCFL), etc.~\citep{Buballa:2003qv}. However, they are out of the scope of this article.

We performed calculations of the thermal evolution of NSs modeled within the IST EOS as this model provides an agreement with HESS J1731-347 mass-radius measurements within 1$\sigma$ CI. The IST EOS is formulated in terms of nucleons characterized by an effective hard-core radius yielding a short-range repulsion between them. The latter was fixed from the fit of heavy-ion collision data~\citep{Sagun:2017eye}, while the IST contribution was implemented by accounting for an interparticle interaction at high density. The model was applied to describe the nuclear liquid-gas phase transition and its critical point~\citep{Sagun:2016nlv}, proton flow constraint~\citep{Ivanytskyi:2017pkt}, and further generalized to describe NSs showing a big application range of the unified IST approach~\citep{Sagun:2018sps, Sagun:2018cpi}. The considered parameterization gives the values of the symmetry energy $E_\mathrm{sym} = 30.0$ MeV, symmetry energy slope $L = 93.2$ MeV and nuclear incompressibility factor $K_{0} = 201.0$ MeV at the normal nuclear density~\citep{NSOscillationsEOS}. For realistic modeling of the outer layers, the IST EOS is supplemented by the Haensel-Zdunik (HZ) EOS for the outer crust and the Negele-Vautherin (NV) EOS for the inner crust~\citep{1990A&A...227..431H,Negele:1971vb}.

The pairing of nucleons in simulations of the thermal evolution of NSs depends on the pairing channel and the considered gap model. By adopting the thermal evolution code described in~\citet{2020A&A...642A..42S} we found that the best agreement with HESS J1731-347 data is obtained for $^{1}S_{0}$ neutron and proton pairing, in the inner crust and core, respectively, described by the SFB~\citep{Schwenk:2002fq} and CCDK~\citep{Chen:1993bam} models, as well as $^{3}P_{2}$ pairing of neutrons \citep{Shternin:2010qi}. These results are very much in line with predictions of other works~\citep{PhysRevC.87.015804, Negreiros_2018, PhysRevC.85.035805, PhysRevC.107.025806}. The results are presented on the upper panel of Fig.~\ref{fig:cooling}).

We also analyze the effect of different envelope compositions: a hydrogen-rich envelope that contains the fraction of light elements $\eta=\Delta M/M= 10^{-7}$ (depicted by solid curves in Fig.~\ref{fig:cooling}) and one containing more heavy elements (dashed curves in Fig.~\ref{fig:cooling}). Here $\Delta M$ is the mass of light elements in the upper envelope. 

The cooling simulations presented on the lower panel of Fig.~\ref{fig:cooling} were performed for HSs described within the DD2npY-T EOS (hadron phase) and RDF approach (quark phase). Modeling of the thermal evolution of HSs incorporated 2SC pairing between quarks obtained in a self-consistent calculation within the RDF model. Moreover, for the inner and outer crusts, we adopted the same HZ and NV EOSs~\citep{1990A&A...227..431H,Negele:1971vb} as for the NS case.

As the 2SC pairing yields a very good agreement with the HESS J1731-347 measured surface temperature, we do not see the necessity to include the CFL phase, as it will cause even stronger neutrino suppression providing an equivalently good data fit. 
 
The observational data were taken from~\citet{Beznogov:2014yia}. We consider 2$\sigma$ error bars for the available data, otherwise a factor of 0.5 and 2 in both the temperature and the age, excluding the upper limits. The sources are 0 - CasA NS, 1 - PSR J0205+6449 (in 3C58), 2 - PSR B0531+21 (Crab), 3 - PSR J1119-6127, 4 - RX J0822-4300 (in PupA), 5 - PSR J1357-6429, 6 - PSR B1706-44, 7 - PSR B0833-45 (Vela), 9 - PSR J0538+2817, 10 - PSR B2334+61, 11 - PSR B0656+14, 12 - PSR B0633+1748 (Geminga), 13 - PSR J1741-2054, 14 - RX J1856.4-3754, 15 - PSR J0357+3205 (Morla), 16 - PSR B1055-52, 17 - PSR J2043+2740, 18 - RX J0720.4-3125. The object 8 - XMMU J1731-347 in~\citet{Beznogov:2014yia} was substituted by the HESS J1731-347~\citep{Doroshenko:2022}. An updated data shows a slight increase in the surface temperature. 
\section{Two-fluid approach}
\label{app:DM}

In this work, we consider the DM component as a relativistic Fermi gas composed of noninteracting massive particles possessing a spin of one-half. The corresponding EOS has been extensively studied in the literature, e.g. by~\citet{Nelson:2018xtr,Ivanytskyi:2019wxd,Sagun:2021oml}. 

Following the constraint from the Bullet Cluster~\citep{Clowe_2006} on the negligible cross section between BM and DM, we assume two components to interact only through gravity. As a result, stress-energy tensors of two fluids (i=DM,BM) are conserved separately leading to two coupled Tolman-Oppenheimer-Volkoff (TOV) equations~\citep{PhysRev.55.364,PhysRev.55.374}
\begin{equation}\label{TOV}
\frac{dp_i}{dr}=-\frac{(\epsilon_i +p_i)(M_\mathrm{tot}+4\pi r^3p_\mathrm{tot})}{r^2\left(1-{2M_\mathrm{tot}}/{r}\right)},
\end{equation}
where $M_\mathrm{tot}=M_\mathrm{DM}+M_\mathrm{BM}$ and $p_\mathrm{tot}=p_\mathrm{DM}+p_\mathrm{BM}$ are the total gravitational mass and pressure, respectively. Since it is a two-fluid system we define the value of the central density for each component. After the integration of the TOV equations, we get the gravitational masses of each of the components. Using these gravitational masses, the DM fraction can be expressed as $f_{DM} = \frac{M_\mathrm{DM}}{M_\mathrm{tot}}$. By adjusting the central energy densities of each component we are able to obtain different scenarios of admixed stars, and, in particular, stars with different DM fractions. As was shown by~\citet{Ivanytskyi:2019wxd} the chemical potentials of two components are related to each other as 
\begin{equation}
    \frac{d \ln \mu_B}{dr}=\frac{d \ln \mu_{DM}}{dr} = -\frac{M_\mathrm{tot}+4\pi r^3 p_\mathrm{tot}}{r^2(1-2M_\mathrm{tot}/r)}.
\end{equation}

\bibliography{references}{}
\bibliographystyle{aasjournal}

\end{document}